\documentclass[aps,prl,twocolumn,groupedaddress,showpacs]{revtex4}
\usepackage{graphicx}
\usepackage[usenames]{color}
\usepackage{psfrag}

\newsavebox{\tempbox}

\begin{document}

\title{Measuring Energy Differences by BEC Interferometry
on a Chip}

\author{Florian Baumg{\"a}rtner}
\author{R. J. Sewell}
\altaffiliation[Current address:]{ICFO-Institut de Ciencies Fotoniques, 08860 Castelldefels (Barcelona), Spain}
\author{S. Eriksson}
\altaffiliation[Permanent address:]{Department of Physics, Swansea University, Singleton Park, Swansea SA2 8PP, UK}
\author{I. Llorente-Garcia}
\author{Jos Dingjan}
\author{J. P. Cotter}
\author{E. A. Hinds}
\email[]{ed.hinds@imperial.ac.uk}
\affiliation{Centre for Cold Matter, Blackett Laboratory, Imperial College, Prince Consort Road, London SW7 2AZ, United Kingdom}

\date{\today}

\begin{abstract}
We investigate the use of a Bose-Einstein condensate trapped on an atom chip for making interferometric measurements of small energy differences. We measure and explain the noise in the energy difference of the split condensates, which derives from statistical noise in the number difference. We also consider systematic errors.  A leading effect is the variation of rf magnetic field in the trap with distance from the wires on the chip surface. This can produce energy differences that are comparable with those due to gravity.
\end{abstract}

\pacs{37.25.+k, 37.10.Gh, 06.20.-f}
\maketitle

The phase of an atomic wavefunction can measure small energy differences~\cite{berman1997}. In the last decade, thermal atom interferometers have made remarkably sensitive measurements of rotations, gravity, atomic polarizability, the fine structure constant, and atom-surface interactions~\cite{cronin2009}. The coherent matter waves of
Bose-Einstein condensates (BECs)~\cite{andrews1997} suggest a natural extension to this interferometry. The BEC is conveniently prepared on an atom chip~\cite{fortagh2007}, where it is trapped in a small volume that offers high spatial resolution and opens entirely new prospects for interferometry. The proximity of the BEC to the chip surface lends itself naturally to the study of atom-surface interactions, both electromagnetic~\cite{carusotto2005,harber2005,obrecht2007,hall2007} and gravitational~\cite{dimopoulos2003,ferrari2006,sorrentino2009}. From the perspective of wider applications, the small size of an atom chip offers the prospect of portable devices~\cite{vanZoest2010}.

Interferometry on atom chips has advanced rapidly in the past few
years~\cite{hinds2001,shin2004,shin2005,wang2005,schumm2005,jo2007,jo2007a,jo2007b}. The early difficulty of achieving a fixed initial phase between the split BECs in a double well potential has been overcome~\cite{schumm2005}. State-dependent potentials allow internal atomic states to become entangled with motional states on a chip~\cite{bohi2009}, and it is now possible even to reach below the shot noise level of sensitivity~\cite{gross2010,riedel2010}.

In this letter we consider a BEC, trapped on an atom chip and coherently split by an rf magnetic field~\cite{zobay2001}. We quantify for the first time both the sensitivity and the absolute accuracy of such an interferometer, providing a systematic assessment of atom chip BEC interferometry as a tool for measuring small energy differences.

\begin{figure}[t!]
\centering
\includegraphics[width = 1.0 \columnwidth]{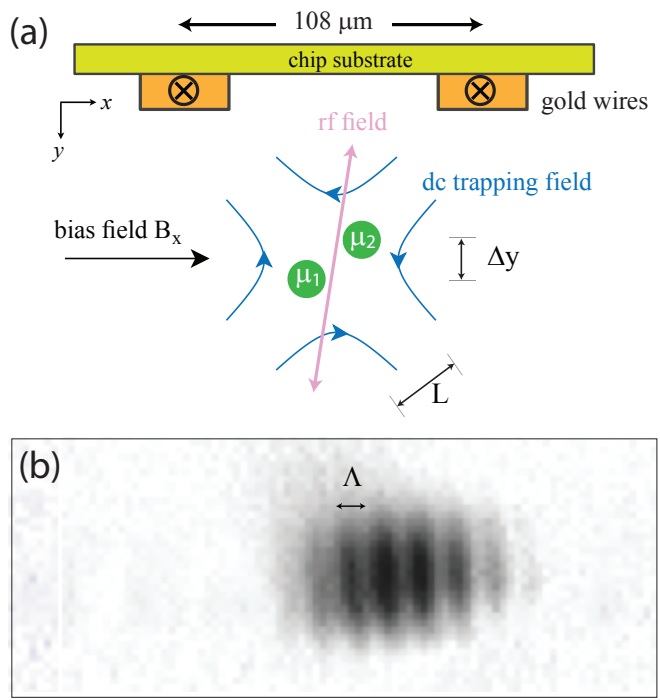}
\caption{Gold wires on the atom chip carry dc and rf currents that, together with a bias field, form a static magnetic trap. The rf field splits the condensate and allows the double trap to be rotated. Gravity acts along the $y$-axis. The atoms are released to interfere in free fall below the trap. A single-shot absorption image is shown.
\label{fig:chip}}
\end{figure}
Our apparatus and the atom chip have been described elsewhere~\cite{sewell2010}. In brief, the magnetic field that constitutes the trap is formed by currents in two parallel Z-shaped gold wires~\cite{reichel2002} separated (centre-to-centre) by 108\,$\mu$m, together with a uniform bias field $B_x$, as in Fig.~\ref{fig:chip}(a). This holds $^{87}$Rb atoms in state \( \left|F=2,m_F=2\right> \) at a distance $y=130\,\mu$m from the chip surface, with radial and axial trap-frequencies of $f_r=2$\,kHz and $f_z=28$\,Hz. An additional magnetic field along $\hat{z}$ adjusts the field magnitude $B_0$ at the potential minimum.

After loading the magnetic trap from a magneto-optical trap, a short rf evaporation ramp produces a nearly pure BEC of $N=1.5\times 10^4$ atoms. The condensate is then split using an rf magnetic field~\cite{schumm2005} produced by currents $180^{\circ}$ out of phase in the two Z-wires, as illustrated in Fig.~\ref{fig:chip}(a). The Larmor frequency at the centre of the dc trap is $g_F\mu_B B_0/h = 630$\,kHz, and the rf frequency is fixed at a detuning of $90$\,kHz below this. The double well is formed by ramping up the rf amplitude over $20\,$ms so that the coupling between states \(\left|m_F=2\right> \) and \(\left|m_F=1\right> \) increases to a typical Rabi frequency of $300$\,kHz. This separates the condensate into two parts that start to accumulate a phase difference according to the difference in their chemical potentials.

After a suitable integration time \(\tau\), we release the condensates and allow them to fall freely for a time $T=12.4$\,ms. The interference fringe pattern in the atomic density is then recorded, as shown in Fig.~\ref{fig:chip}(b), by absorption of resonant light. We fit a modulated Gaussian, \(G(x)\left(1+\alpha\cos\left(\frac{2 \pi x}{\Lambda} + \phi\right) \right) \), to a slice through the centre of the absorption image in order to determine the phase difference \( \phi \). The line of zero phase is the perpendicular bisector of the line joining the condensates, but that is not visible in the camera image. In practice it is adequate to define \(x=0\) as the centre of the  Gaussian \(G(x)\) because this exhibits very little drift or jitter relative to the zero phase line.

The expansion velocities of the atoms are constant (in the centre-of-mass frame and neglecting a very brief initial acceleration) and one therefore expects the fringe spacing \(\Lambda\) to be related to the initial condensate separation \(L\) by $L= hT/m\Lambda$, where $m$ is the mass of the rubidium atom and $h$ is Planck's constant. We have tested this equation experimentally and find that it is indeed accurately followed provided the initial two condensates are not overlapping. We therefore use the period of the fringes to infer the initial cloud splitting in the experiments that follow.

\begin{figure}[b!]
\centering
{\footnotesize
\begin{psfrags}
\def\PFGstripminus-#1{#1}%
\def\PFGshift(#1,#2)#3{\raisebox{#2}[\height][\depth]{\hbox{%
  \ifdim#1<0pt\kern#1 #3\kern\PFGstripminus#1\else\kern#1 #3\kern-#1\fi}}}%
\providecommand{\PFGstyle}{}%
\psfrag{HoldtimeTa}[tc][tc]{\PFGstyle Hold time $\tau$ (ms)}%
\psfrag{Phasesprea}[bc][bc]{\PFGstyle Phase spread $\sigma_{\phi}$ (rad)}%
\psfrag{S0}[tc][tc]{\PFGstyle $0$}%
\psfrag{S12}[tc][tc]{\PFGstyle $10$}%
\psfrag{S21}[tc][tc]{\PFGstyle $2$}%
\psfrag{S41}[tc][tc]{\PFGstyle $4$}%
\psfrag{S61}[tc][tc]{\PFGstyle $6$}%
\psfrag{S81}[tc][tc]{\PFGstyle $8$}%
\psfrag{Sm21}[tc][tc]{\PFGstyle $-2$}%
\psfrag{Sm41}[tc][tc]{\PFGstyle $-4$}%
\psfrag{W0}[cr][cr]{\PFGstyle $0$}%
\psfrag{W11}[cr][cr]{\PFGstyle $1.0$}%
\psfrag{W151}[cr][cr]{\PFGstyle $1.5$}%
\psfrag{W21}[cr][cr]{\PFGstyle $2.0$}%
\psfrag{W5}[cr][cr]{\PFGstyle $0.5$}%
\includegraphics[width = 1.0 \columnwidth]{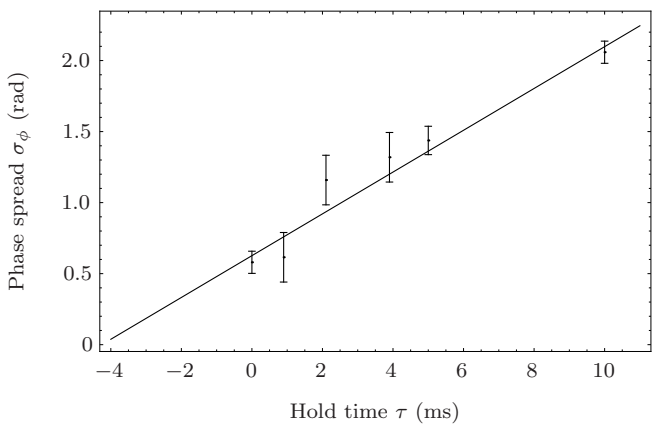}
\end{psfrags}
}
\caption{Standard deviation of the phase of the interference pattern versus hold time \(\tau\) after reaching full splitting. This is measured by many repetitions of the experiment: there were 100 shots each at 0 and 10\,ms, 61 shots at 5\,ms and 20 shots each at 0.9, 2.1, and 3.9\,ms. The solid line is a fit to the data showing $\partial\sigma_{\phi}/\partial\tau=$147(10)\,rad/s.
\label{fig:phase}
}
\end{figure}

First we investigate the relative phase \(\phi\) after splitting the condensate and holding it for a time \(\tau\). Repeated experiments give a distribution of \(\phi\), with a mean value of \(\langle\phi\rangle=\arctan(\langle\sin\phi\rangle/\langle\cos\phi\rangle)\) and a standard deviation of \(\sigma_{\phi}=\sqrt{-2\ln\sqrt{\langle\sin\phi\rangle^2+\langle\cos\phi\rangle^2}}\)~\cite{mardia2000}. As shown in Fig.~\ref{fig:phase} we find that this phase spread grows linearly with the hold time \(\tau\) and a fit to the data gives the slope as $147(10)$rad/s. This growth is driven by the difference \(k=n_1-n_2\) between the numbers of atoms in the two condensates, which causes a difference in the chemical potentials \(\Delta\mu=\mu(n_1)-\mu(n_2)\)~\cite{javanainen1997}. For small fluctuations, the standard deviation of \(\Delta\mu\) is \(\sigma_{\Delta\mu}=\frac{\partial \mu(n)}{\partial n}|_{_{N/2}}\sigma_k\), where \(\sigma_k\) is the standard deviation of \(k\). The corresponding growth in the standard deviation of the phase, \(\sigma_\phi=\frac{1}{\hbar}\sigma_{\Delta\mu}\tau \), is the linear growth that we see. Our measurement therefore indicates that the uncertainty in the chemical potential difference is \(\frac{1}{h}\sigma_{\Delta\mu} = 23(2)\)\,Hz, a value that we explain below. The phase spread seen at \(\tau=0\) is due to this mechanism operating over the previous \(\sim 4\)\,ms as we increase the splitting to \(L\). In the course of this measurement we found that the jitter in the position of the atom cloud produces correlated phase noise, which we have systematically corrected in producing Fig.~\ref{fig:phase}. This correction only affects the earliest times because it amounts to a fixed $0.5$\,rad that adds in quadrature with the growing phase noise. There is also a very small fixed phase uncertainty coming from the Fourier relation \(\sigma_k \sigma_\phi=1\), but this is entirely negligible.

When the interferometer is used to make a measurement, the signal of interest also derives from the difference between the two chemical potentials. The uncertainty after one shot is therefore $23$\,Hz, regardless of the measurement time, provided \(\tau\) is long enough that the initial \(\tau=0\) phase error is small. With \(p\) repetitions the limit of sensitivity becomes $23$\,Hz$/\sqrt{p}$. For comparison, the gravitational potential energy of a \(^{87}\)Rb atom increases by $2$\,Hz when it is lifted through a height of $1$\,nm. We turn next to the measurement of this effect.

The BEC is first split along the \(x\)-direction to a separation $L = 3.47(4)\,\mu$m. The double well is then rotated uniformly over a time $\tau/2$ through an angle $\theta$ around $\hat{z}$ by adjusting the relative magnitudes of the rf currents~\cite{hofferberth2006}. The weight \(mg\) of the atoms causes an increase (decrease) in the potential energy of the BEC that moves up (down). The double well is then tilted back to its original orientation, again uniformly over a time $\tau/2$, resulting in a gravitationally-induced phase shift of $\frac{1}{\hbar}\int_{t=0}^{\tau}m g\delta y(t) dt$, where \(\delta y\)  is the difference in height between the centres of mass of the two condensates. For our small angles \(\theta\), this is well approximated  by \(\frac{1}{2\hbar}mg\Delta y\,\tau\), where \(\Delta y=L\theta\).
\begin{figure}[t!]
\centering
\begin{psfrags}
\def\PFGstripminus-#1{#1}%
\def\PFGshift(#1,#2)#3{\raisebox{#2}[\height][\depth]{\hbox{%
  \ifdim#1<0pt\kern#1 #3\kern\PFGstripminus#1\else\kern#1 #3\kern-#1\fi}}}%
\providecommand{\PFGstyle}{}%
\psfrag{a}[tc][tc]{\PFGstyle {\large (a)} }%
\psfrag{b}[tc][tc]{\PFGstyle {\large (b)} }%
\psfrag{S0}[tc][tc]{\PFGstyle $0$}%
\psfrag{S13}[tc][tc]{\PFGstyle $100$}%
\psfrag{S153}[tc][tc]{\PFGstyle $150$}%
\psfrag{S23}[tc][tc]{\PFGstyle $200$}%
\psfrag{S52}[tc][tc]{\PFGstyle $50$}%
\psfrag{Sm13}[tc][tc]{\PFGstyle $-100$}%
\psfrag{Sm153}[tc][tc]{\PFGstyle $-150$}%
\psfrag{Sm23}[tc][tc]{\PFGstyle $-200$}%
\psfrag{Sm52}[tc][tc]{\PFGstyle $-50$}%
\psfrag{W0}[cr][cr]{\PFGstyle $0$}%
\psfrag{W12}[cr][cr]{\PFGstyle $10$}%
\psfrag{W152}[cr][cr]{\PFGstyle $15$}%
\psfrag{W21}[cr][cr]{\PFGstyle $2$}%
\psfrag{W22}[cr][cr]{\PFGstyle $20$}%
\psfrag{W252}[cr][cr]{\PFGstyle $25$}%
\psfrag{W41}[cr][cr]{\PFGstyle $4$}%
\psfrag{W51}[cr][cr]{\PFGstyle $5$}%
\psfrag{W61}[cr][cr]{\PFGstyle $6$}%
\psfrag{Wm21}[cr][cr]{\PFGstyle $-2$}%
\psfrag{Wm41}[cr][cr]{\PFGstyle $-4$}%
\psfrag{Wm51}[cr][cr]{\PFGstyle $-5$}%
\psfrag{Wm61}[cr][cr]{\PFGstyle $-6$}%
\psfrag{x0}[tc][tc]{\PFGstyle $0$}%
\psfrag{x11}[tc][tc]{\PFGstyle $1$}%
\psfrag{x1252}[tc][tc]{\PFGstyle $12.5$}%
\psfrag{x12}[tc][tc]{\PFGstyle $10$}%
\psfrag{x152}[tc][tc]{\PFGstyle $15$}%
\psfrag{x1752}[tc][tc]{\PFGstyle $17.5$}%
\psfrag{x22}[tc][tc]{\PFGstyle $20$}%
\psfrag{x251}[tc][tc]{\PFGstyle $2.5$}%
\psfrag{x2}[tc][tc]{\PFGstyle $0.2$}%
\psfrag{x4}[tc][tc]{\PFGstyle $0.4$}%
\psfrag{x51}[tc][tc]{\PFGstyle $5$}%
\psfrag{x6}[tc][tc]{\PFGstyle $0.6$}%
\psfrag{x751}[tc][tc]{\PFGstyle $7.5$}%
\psfrag{x8}[tc][tc]{\PFGstyle $0.8$}%
\psfrag{x}[Bc][Bc]{\PFGstyle $\Delta y$ (nm)}%
\psfrag{y0}[cr][cr]{\PFGstyle $0$}%
\psfrag{y11}[cr][cr]{\PFGstyle $1$}%
\psfrag{y1252}[cr][cr]{\PFGstyle $12.5$}%
\psfrag{y12}[cr][cr]{\PFGstyle $10$}%
\psfrag{y152}[cr][cr]{\PFGstyle $15$}%
\psfrag{y1752}[cr][cr]{\PFGstyle $17.5$}%
\psfrag{y1}[bc][bc]{\PFGstyle $\phi$\,(rad)}%
\psfrag{y22}[cr][cr]{\PFGstyle $20$}%
\psfrag{y251}[cr][cr]{\PFGstyle $2.5$}%
\psfrag{y2A}[cr][cr]{\PFGstyle $0.2$}%
\psfrag{y2}[bc][bc]{\PFGstyle $\phi/(\tau/2)$\,(rad/ms)}%
\psfrag{y4}[cr][cr]{\PFGstyle $0.4$}%
\psfrag{y51}[cr][cr]{\PFGstyle $5$}%
\psfrag{y6}[cr][cr]{\PFGstyle $0.6$}%
\psfrag{y751}[cr][cr]{\PFGstyle $7.5$}%
\psfrag{y8}[cr][cr]{\PFGstyle $0.8$}%
\psfrag{zA}[Bc][Bc]{\PFGstyle $~$}%
\psfrag{z}[bc][bc]{\PFGstyle $~$}%
\psfrag{n1}[tc][tc]{\PFGstyle \color{Green}{$4.0$\,ms}}%
\psfrag{n2}[tc][tc]{\PFGstyle \color{Blue}{$3.2$\,ms}}%
\psfrag{n3}[tc][tc]{\PFGstyle \color{BrickRed}{$2.0$\,ms}}%
\includegraphics[width=1.0 \columnwidth]{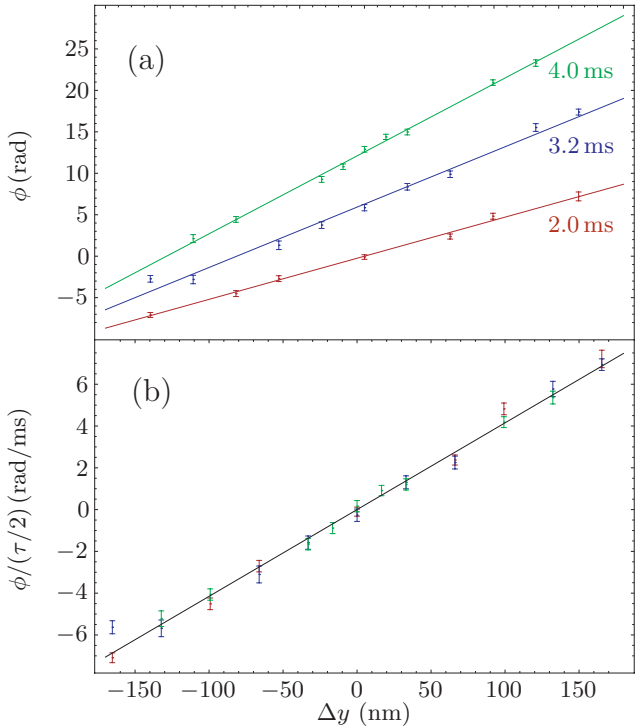}
\end{psfrags}
\caption{(a) Interferometer phase \(\phi\) versus height difference \(\Delta y\) between the two condensates for three measurement times $\tau=2$\,ms \mbox{, } $3.2$\,ms \mbox{ and } $4$\,ms. The lines are offset for clarity. (b) On dividing by \(\tau/2\) to obtain the energy gradient, the points all lie on a universal curve of slope $42.56(72)$\,rad\,s$^{-1}$/nm or $6.61(11)$Hz/nm. The solid lines are linear best fits to the data.}
\label{fig:interferometry}
\end{figure}

Figure~\ref{fig:interferometry}(a) shows three sets of interferometer phases (offset for clarity), measured as a function of the maximum height difference $\Delta y$. Each point is the mean phase \(\langle\phi\rangle\) given by $20$ repeated measurements and the error bars indicate the standard error. The slopes of these lines differ because they are measured with three different interaction times: $\tau=2$\,ms \mbox{, } $3.2$\,ms \mbox{ and } $4$\,ms. When each series is divided by its corresponding value of \(\tau/2\) the whole data set lies on a single line, shown in Fig.~\ref{fig:interferometry}(b). The slope of this line divided by $2\pi$ is $\frac{1}{2\pi}\frac{2}{\tau}\frac{\partial\phi}{\partial\Delta y} = 6.61(10)$\,Hz/nm. The measurement thus demonstrates a sensitivity of $100$\,mHz/nm, but the interferometer is measuring more than gravity.

\begin{figure}[t]
\centering
\begin{psfrags}
\psfrag{820}[tc][tc]{$820$}%
\psfrag{830}[tc][tc]{$830$}%
\psfrag{840}[tc][tc]{$840$}%
\psfrag{850}[tc][tc]{$850$}%
\psfrag{860}[tc][tc]{$860$}%
\psfrag{0.0}[tc][tl]{$0.0$}%
\psfrag{0.5}[tc][tl]{$0.5$}%
\psfrag{1.0}[tc][tl]{$1.0$}%
\psfrag{1.5}[tc][tl]{$1.5$}%
\psfrag{2.0}[tc][tl]{$2.0$}%
\psfrag{2.5}[tc][tl]{$2.5$}%
\psfrag{X}[tc][tc]{$\nu_{\mathrm{rf}}$ (kHz)}%
\psfrag{Y}[tb][tc]{atom density (arb. un.)}%
\psfrag{a}[cr][cc]{{\footnotesize $-145$\,nm}}%
\psfrag{b}[cr][cc]{{\footnotesize$-53$\,nm}}%
\psfrag{c}[cr][cc]{\footnotesize{$11$\,nm}}%
\psfrag{d}[tr][tc]{{\footnotesize$187$\,nm}}%
\psfrag{0}[tc][tl]{{\tiny$0$}}%
\psfrag{200}[tc][tl]{{\tiny$200$}}%
\psfrag{400}[tc][tl]{{\tiny$400$}}%
\psfrag{600}[tc][tl]{{\tiny$600$}}%
\psfrag{800}[tc][tl]{{\tiny$800$}}%
\psfrag{150}[tc][tl]{{\tiny$~$}}%
\psfrag{101}[tc][tl]{{\tiny$-100$}}%
\psfrag{100}[tc][tl]{{\tiny$100$}}%
\psfrag{50}[tc][tl]{{\tiny$~$}}%
\psfrag{-50}[tc][tl]{{\tiny$~$}}%
\psfrag{-100}[tc][tl]{{\tiny$-100$}}%
\psfrag{-}[tc][tl]{{\tiny$~$}}%
\psfrag{dn}[tb][tc]{{\tiny $\Delta \nu_{0}$\,(Hz)}}%
\psfrag{dy}[tc][tc]{{\tiny $\Delta y$\,(nm)}}%
\includegraphics[width= 1.0 \columnwidth]{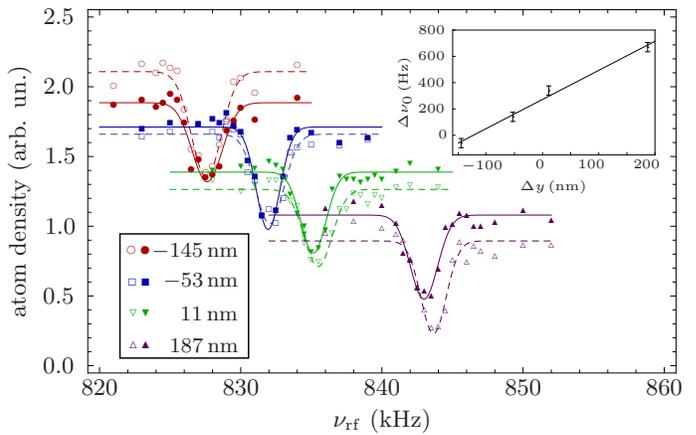}
\end{psfrags}
\caption{Spectra of atom loss from the double well potential versus rf frequency, measured at four different values of the height difference \(\Delta y\). Open (filled) symbols indicate atom densities for the left (right) condensate. The data have been offset for clarity, and lines show the best fits to a gaussian profile. Inset: measured dependence of the frequency shift \(\Delta \nu_{0}\) on \(\Delta y\).
\label{fig:spectroscopy}}
\end{figure}

Introducing a height difference causes the rf magnetic field at one condensate to increase because of its altered distance from the chip, while the other decreases. The resulting magnetic energy difference \cite{vanEs2008} must be subtracted before we can determine the weight of the atoms.  An \(F=2\) atom interacting through its magnetic dipole moment with combined static and rf magnetic fields has five equally spaced eigenenergies~\cite{haroche1971} (the coupling to \(F=1\) is negligible).  Thus, the magnetic energy \(W_{\mathrm{mag}}\) of our trapped (nominally) \(m_F=2\) atoms is just twice the splitting \(h\nu_{0}\) between adjacent levels. We determine the magnetic energy difference \(\Delta W_{\mathrm{mag}}\) by measuring the frequency shift \(\Delta\nu_{0}\) between the values of \(\nu_{0}\) for the left and right condensates. The condensate is split and rotated to produce the height difference \(\Delta y\), then held for $30$\,ms while a weak rf probe field drives spin-flip transitions that deplete the number of trapped atoms~\cite{easwaran2010}. In order to detect the two condensates separately we switch off all the rf fields, leaving them displaced on opposite sides of the static magnetic trap. The clouds then oscillate, periodically converting their initial difference of position into a difference of momentum. We switch off the dc magnetic trap at a moment of maximum momentum difference, which allows us fully to resolve the clouds by absorption imaging after $3$\,ms of free fall. Figure~\ref{fig:spectroscopy} shows the atom density in each condensate versus the frequency of the rf probe for several values of \(\Delta y\). Fits to these line profiles determine the shift \(\Delta\nu_{0}\), which is plotted against \(\Delta y\) in the inset to Fig.~\ref{fig:spectroscopy}. On doubling the slope of this graph, we find that \(\frac{1}{h}\frac{\Delta W_{\mathrm{mag}}}{\Delta y}=4.44(30)\)\,Hz/nm. This is consistent with the rf field gradient we expect from the geometry of our chip. There is also a change of trap frequency with height, which causes a difference in the energies due to the atom-atom repulsion, but we calculate that the systematic error due to this is almost a thousand times smaller than \(\Delta W_{\mathrm{mag}}\) and therefore negligible.

On subtracting the magnetic contribution from the total interferometer phase shift, we obtain the expected gravitational shift of \(2.17(32)\)\,Hz/nm, where the error bar is now large because it is dominated by uncertainty in the magnetic correction. Other systematic uncertainties are readily controlled to better than \(1\%\). These include the magnification of the optical imaging system, the determination of the splitting distance \(L\) and the calibration of the tilt angle.

Two primary areas of improvement suggest themselves. First, the correction for magnetic field difference can be much reduced  by using a symmetrical geometry; for example mounting the chip vertically and splitting parallel to the surface for a gravity measurement. The rf field gradients can also be made quadratically smaller by moving the rf wires further away. With these improvements, we anticipate that absolute measurements should be possible at better than \(1\%\) accuracy as outlined above. Second, the energy resolution can be improved. In an interferometer where the atoms make a random choice between the left and right wells, the noise in the number difference is just the shot noise \(\sigma_k=\sqrt{N}\), amounting to \(0.8\%\) of the atoms in our case. Making the Thomas-Fermi approximation, we then obtain a simple expression for the fluctuation of the chemical potential difference:  \(\sigma_{\Delta\mu}=\hbar(\frac{72}{125 }\frac{m}{\hbar })^{1/5}\omega^{6/5}a_s^{2/5}N^{-1/10}\), where \(\omega\) is the geometric mean frequency of either trap after splitting. For our experiment, where \(\omega=2\pi\times 360\)\,Hz, this formula gives \(\frac{1}{h}\sigma_{\Delta\mu}=19\)\,Hz, as compared with the measured $23$\,Hz, showing that our interferometer is operating close to its ideal sensitivity. The energy resolution could be enhanced by lowering \(\omega\), but this has limited scope since the trap must remain strong enough to support the atoms against gravity. Alternatively, it is possible to squeeze the number difference, which has been shown to reduce the energy fluctuations in sodium by a factor of 10~\cite{jo2007a}. Finally, it is possible to reduce interactions by a Feshbach resonance, as demonstrated, for example, in Ref~~\cite{fattori2008} using $^{39}$K. With some combination of these measures one can expect a fundamental noise level $1\mbox{Hz}/\sqrt{p}$ or below, depending on the choice of atom. However, this high sensitivity does not translate easily into high absolute measurement precision because of systematic errors appearing at the $1\,\%$ level. Of these, the most troublesome is the change in energy of the trap bottom due to spatial variation of the rf field strength.

The small size of the atom cloud and its proximity to a surface make trapped BEC interferometry attractive for mapping atom-surface interactions. For example, a Rb atom $1\,\mu$m from a plane conductor has a Casimir-Polder interaction energy of $270\,\mbox{Hz}$~\cite{hinds1991}, decreasing to $3.3\,\mbox{Hz}$ at a distance of $3\,\mu$m. Over this range, it should be possible to make measurements with $\sim 1\%$ accuracy, limited at large distance by the noise level and at short distance by uncertainty in the spatial distribution of the atoms. This offers the possibility of improving over the existing measurements of the Casimir-Polder interaction~\cite{sukenik1993},\cite{harber2005} and of its temperature dependence~\cite{obrecht2007}. Also promising is the atomic Bloch oscillation method, as discussed in \cite{ferrari2006}. The related phenomenon of Casimir attraction between two macroscopic bodies has been measured with 15\% accuracy for parallel plates~\cite{bressi2002} and with precision in the range $1-5\%$ between plane and curved surfaces~\cite{casimir}.

In summary, we have investigated the atom-chip BEC interferometer as a way to measure small energy differences. The fundamental resolution is determined by fluctuations of the chemical potential difference resulting from the uncertainty in atom number difference. In an rf double-well interferometer, the achievable $1\,\mbox{Hz}$ noise level is small enough to permit detection of a Rb atom moving through less than $1\,\mbox{nm}$ in the earth's gravity. We note that this type of interferometer is suitable for accurate measurements of atom-surface interactions.

\begin{acknowledgments}
This work was supported by the UK EPSRC, by the Royal Society, and by the European Commission through the SCALA and AtomChips networks. We are indebted to Jon Dyne for many technical contributions.
\end{acknowledgments}


\end{document}